\documentclass[sigconf, pbalance]{acmart}

\usepackage{multirow}
\usepackage{siunitx}

\usepackage[utf8]{inputenc}
\usepackage[dvipsnames]{xcolor}
\usepackage{xspace}
\usepackage[linesnumbered,ruled,vlined]{algorithm2e}
\usepackage{mathtools}

\usepackage{colortbl}
\usepackage{pifont}

\usepackage{bbding}

\newcommand{\cmark}{\ding{51}}
\newcommand{\xmark}{\ding{55}}

\usepackage{listings}
\usepackage{url}
\usepackage{subcaption}

\usepackage{enumitem}
\setlist{nosep, leftmargin=*, topsep=0pt, partopsep=0pt, parsep=0pt, itemsep=0pt}

\definecolor{codegreen}{rgb}{0,0.6,0}
\definecolor{codegray}{rgb}{0.5,0.5,0.5}
\definecolor{codepurple}{rgb}{0.58,0,0.82}
\definecolor{backcolour}{rgb}{0.95,0.95,0.92}

\newcommand{\llm}{{\sc llm}}

\newcommand{\ie}{\textit{i.e.,}}
\newcommand{\eg}{\textit{e.g.,}}
\newcommand{\etal}{\textit{et al.}}

\newcommand{\gpu}{{\sc GPU}}

\lstdefinestyle{mystyle}{
    backgroundcolor=\color{backcolour},   
    commentstyle=\color{codegreen},
    keywordstyle=\color{magenta},
    numberstyle=\tiny\color{codegray},
    stringstyle=\color{codepurple},
    basicstyle=\ttfamily\footnotesize,
    breakatwhitespace=false,         
    breaklines=true,                 
    captionpos=b,                    
    keepspaces=true,                 
    numbers=left,                    
    numbersep=5pt,                  
    showspaces=false,                
    showstringspaces=false,
    showtabs=false,                  
    tabsize=2
}

\copyrightyear{2026}
\acmYear{2026}
\setcopyright{cc}
\setcctype{by}
\acmConference[ICSE '26]{2026 IEEE/ACM 48th International Conference on Software Engineering}{April 12--18, 2026}{Rio de Janeiro, Brazil}
\acmBooktitle{2026 IEEE/ACM 48th International Conference on Software Engineering (ICSE '26), April 12--18, 2026, Rio de Janeiro, Brazil}
\acmPrice{}
\acmDOI{10.1145/3744916.3787793}
\acmISBN{979-8-4007-2025-3/2026/04}

\AtBeginDocument{%
  }

\begin{document}
\raggedbottom

\title{\emph{FlipFlop}: A Static Analysis-based Energy Optimization Framework for GPU Kernels}

\author{Saurabhsingh Rajput}
\affiliation{
  \institution{Dalhousie University}
  \city{Halifax}
  \country{Canada}}
\email{saurabh@dal.ca}

\author{Alex Brandt}
\affiliation{
  \institution{Dalhousie University}
  \city{Halifax}
  \country{Canada}}
\email{ABrandt@dal.ca}

\author{Vadim Elisseev}
\affiliation{%
  \institution{IBM Research}
  \city{Warrington}
  \country{UK}}
\email{vadim.v.elisseev@ibm.com}

\author{Tushar Sharma}
\affiliation{
  \institution{Dalhousie University}
  \city{Halifax}
  \country{Canada}}
\email{tushar@dal.ca}

\renewcommand{\shortauthors}{Rajput \etal{}}

\begin{abstract}
Artificial Intelligence (AI) applications, such as Large Language Models, are primarily driven and executed by Graphics Processing Units (\gpu{}s).  
These GPU programs (kernels) consume substantial amounts of energy, yet software developers often lack the hardware expertise and ad hoc knowledge required to optimize for power efficiency. We propose \emph{\textbf{FlipFlop}}, a framework using static code analysis to predict energy consumption and recommend Pareto-optimal thread block configurations considering both power consumption and execution time. Our framework requires no runtime execution and analyzes PTX code, a low-level instruction set for CUDA-enabled \gpu{}s. It is validated across a diverse set of \gpu{}s and kernels, including multi-head attention, convolution, and matrix multiplication. \emph{FlipFlop} achieves 83\% accuracy in identifying locally optimal energy-efficient configurations, while also minimizing developer effort by reducing the optimization search space by 93.4\%. For multi-head attention kernels, it yields up to 79\% energy savings and 106\% throughput gains relative to NVIDIA's occupancy heuristic. By integrating static analysis with real-time monitoring and providing explainable optimization guidance, \emph{FlipFlop} empowers developers to create sustainable, high-performance \gpu{} software which minimizes environmental and computational costs.
\end{abstract}


\maketitle

\section{Introduction}
\label{sec:introduction}

The rapid evolution of Artificial Intelligence (AI), particularly Large Language Models (LLMs)~\cite{de2023growing,chien2023reducing,samsi2023wordswattsbenchmarkingenergy}, has led to the emergence of AI-enabled software systems with unprecedented computational demands.
Training and deploying these models consumes significant energy; the International Energy Agency notes that data center electricity---driven heavily by AI training and inference---accounts for 2\% of global electricity, projected to more than double by 2026, surpassing Canada's national power consumption~\cite{jevonsSasha2025,guidi2024environmentalburdenunitedstates,ieaDatacenters}.

Traditional software engineering (SE) practices rely on programming abstraction layers that conceal underlying complexity. Frameworks like TensorFlow~\cite{tensorflowTensorFlow} and PyTorch cover hardware complexity, making AI development easier. While these abstractions provide usability, hardware-agnostic execution, and accelerated development, they create \emph{hardware opacity} that prevents developers from leveraging architecture-specific optimizations that could dramatically reduce energy consumption.

This hardware opacity manifests concretely in common development scenarios. CUDA kernels---programs executed on NVIDIA GPUs---are critical for optimizing AI workloads but require manual tuning for efficiency~\cite{nvidiaCuda2025}. Consider developing custom attention mechanisms or optimizing scientific computing workflows: developers could achieve substantial energy savings by manually tuning CUDA kernels to optimize memory access patterns. However, the specialized expertise required creates a significant barrier; most developers default to framework-provided implementations prioritizing convenience over energy efficiency. This efficiency gap compounds at scale: inefficiencies across months-long model training and millions of inference calls lead to substantial wastage in energy and time.
This problem is aggravated by AI-generated code. Current LLMs are predominantly trained on open-source repositories lacking rigorous efficiency checks~\cite{vartziotis2024learncode}. When these models generate new code, they replicate and amplify existing inefficiencies. As this AI-generated code enters public codebases and becomes training data for future models, we risk entering a degenerative cycle known as ``\textit{Model Collapse}''~\cite{modelCollapse}, where progressively inefficient implementations generate even less optimized code.

\textbf{Addressing these systemic inefficiencies requires new approaches that integrate hardware-aware optimizations into SE workflows, enabling developers to prioritize energy efficiency without deep hardware expertise.} Specialized frameworks such as CUDA~\cite{nvidiaCuda2025}, ROCm~\cite{rocm2025}, and SYCL~\cite{sycl2025} enable direct hardware programming, but optimizing their parameters---particularly thread block configurations---remains a manual, expertise-intensive bottleneck. Each kernel must be hand-tuned by developers with specialized hardware knowledge, creating critical challenges in large-scale AI deployment pipelines. This challenge is exacerbated by heterogeneous hardware environments where modern AI systems span multiple GPU architectures (\eg{} mixing NVIDIA's Ampere, Hopper, and Blackwell generations), requiring per-device tuning that multiplies engineering effort.

Current approaches to automating this tuning process remain fundamentally constrained by runtime measurement reliance. State-of-the-art tools such as Kernel Tuner~\cite{kerneltuner} require \textit{execution} to measure energy usage, making comprehensive optimization impractical at LLM scales. For an AI workload with $n$ configuration parameters, exhaustive profiling requires $O(2^n)$ runs---wasting significant energy and time. This runtime dependency creates three prohibitive costs: (1) \textit{profiling overhead} from executing hundreds of kernel variants; (2) \textit{energy waste} as suboptimal configurations compound across millions of tokens; and (3) \textit{cold-start delays} when transitioning between GPU architectures (\eg{} migrating from NVIDIA Ampere to Blackwell GPUs requires repeating optimization from scratch). These challenges are compounded by restricted hardware access in cloud environments and shared clusters, where power monitoring often requires privileged access. Consequently, energy inefficiencies typically surface only post-deployment, necessitating costly re-engineering cycles.
Critically, preliminary energy assessment, essential for long-running training jobs, remains practically infeasible with current runtime-dependent approaches. 

These limitations highlight the need for \textit{static energy modeling} techniques that analyze and optimize code without execution. Desirable features include: (1) identifying energy hotspots in source code during development rather than deployment, (2) ranking GPU configurations by efficiency before runtime testing, and (3) providing explainable optimization guidance to developers. 

In response, we propose \textit{\textbf{FlipFlop}}---a static analysis-based kernel optimization framework for energy-conscious GPU software development. \textit{FlipFlop} analyzes NVIDIA's Parallel Thread Execution (PTX) intermediate representation~\cite{ptx_wiki_2025, nvidiaIntroductionx2014} of GPU kernels to extract memory access patterns, control flow characteristics, and instruction mix without kernel execution. These static features, combined with a calibrated hybrid performance-power model, predict energy-efficient thread block configurations and power limits. The framework enables developers to quickly identify optimal configurations for compute-intensive workloads. Unlike runtime-dependent approaches, our method provides energy-efficient comparisons through static parsing while delivering explainable hardware-aware guidance---advantages that black-box AI techniques typically lack.

Experimental results on multi-head attention (MHA) kernels show that \textbf{\textit{FlipFlop} reduces up to 79\%  energy consumption} per token compared to NVIDIA’s static occupancy-based heuristics \textbf{while achieving up to 106\% throughput gains}, maintaining strict quality-of-service constraints. These findings, validated through a real-world case study with CodeLlama, highlight \textit{FlipFlop}’s ability to optimize code configurations for production LLMs and demonstrate practical benefits for AI-enabled SE workflows.

The study makes the following key \textbf{contributions}.
\begin{itemize}
    \item A lightweight static analysis framework that predicts energy-efficient GPU kernel \textbf{configurations (configs)}  without exhaustive runtime execution, reducing significant profiling overhead.
    \item A hybrid performance-power model integrating PTX-level code analysis with hardware calibration to recommend optimal thread block shapes and power limits.
    \item Explainable optimization guidance for developers, addressing memory access efficiency 
    and power scaling challenges in LLM inference kernels.
    \item Validation on MHA kernels and a Code Llama case study, achieving up to 79\% energy savings and 106\% throughput gains in production settings.
\end{itemize}

\noindent
We have made \textbf{replication package} publicly available~\cite{flipflop_2025}.

\vspace{-2mm}
\section{Background \& Related Work}
\label{sec:background}


\paragraph{\textbf{Transformer Architectures \& LLM Computation}}
LLMs like LLAMA-3~\cite{grattafiori2024llama} utilize Multi-Head Attention (MHA) to capture dependencies via projected matrix multiplications~\cite{vaswani2017attention}. Scaling quadratically with sequence length, MHA requires high bandwidth. Furthermore, autoregressive decoding creates irregular, memory-bound workloads, making these kernels critical for both performance and energy efficiency.

\vspace{-1mm}
\paragraph{\textbf{GPU Architecture \& CUDA Programming Model}}

Modern GPUs for deep learning feature thousands of cores organized into Streaming Multiprocessors (SMs)~\cite{modalWhatStreaming}, each with compute units (FP32/FP64 cores, integer ALUs, tensor cores), a memory hierarchy (registers, shared memory, L1/L2 caches, global memory), and warp schedulers managing instruction dispatch~\cite{gpuGlossary}. CUDA organizes execution into threads, warps (32 threads executing in lockstep), thread blocks (groups of warps sharing SM resources), and grids (sets of blocks). A \textit{kernel}---a parallel function executing across these units---implements core computational routines closer to the specialized hardware~\cite{computeKernelWiki}. Thread block config including both size (total threads) and shape (dimensional arrangement---\eg{} 32×4 vs 16×8), impact memory coalescing (combining adjacent threads' memory accesses into fewer transactions), SM occupancy (active warps per SM), and power draw. CUDA enables developers to implement custom high-performance operations that frameworks such as PyTorch cannot automatically generate. However, this flexibility necessitates careful kernel tuning to ensure hardware resources are utilized optimally, as suboptimal thread block shapes degrades performance~\cite{nvidiacudaguide2025}---creating multidimensional energy optimization challenges that demand specialized solutions.

\vspace{-1mm}
\paragraph{\textbf{Performance \& Power Modeling}}
\label{subsec:modeling}

LLM computational demands grow exponentially with parameter count, while hardware power constraints remain fixed~\cite{luccioni2024power}. Model-level optimizations like quantization, pruning, and FlashAttention reduce complexity and data movement~\cite{zhu2024survey, dao2022flashattention}, while hardware techniques include predictive power modeling~\cite{alavani_prediction_2023}, adaptive power capping~\cite{zamani_saou_2020, you2023zeus}.
However, traditional \gpu{} optimization approaches suffer from three key limitations.
First, simplified performance models such as NVIDIA's Occupancy Calculator~\cite{nvidiaNsightCompute} prioritize thread parallelism while overlooking power dynamics and memory hierarchy effects, affecting energy efficiency. Crucially, these tools require runtime measurements for accurate recommendations and fail to account for thread block shape effects. Second, post-hoc profiling methods, such as Boughzala et al.’s~\cite{BoughzalaPredicting2020}, rely on simplified simulations that fail to capture complex software-hardware interactions, limiting generalizability across workloads.
Third, while existing analytical models such as GPUWattch~\cite{leng_gpuwattch_2013} and the integrated power-performance model of ~\citet{hong_integrated_2010} provide cycle-level power estimates, their assumptions limit their accuracy and applicability on modern, heterogeneous workloads. These performance models and simulations typically ignore the importance of thread block shapes (\ie{} kernel launch parameters or thread block configs), a factor that significantly influences both energy consumption and performance~\cite{brandt_klaraptor_2019}.
Hong and Kim's memory/compute warp parallelism (MWP/CWP) framework~\cite{hong2009analytical} quantifies how effectively thread warps overlap memory operations or saturate compute pipelines. 
Recent extensions by Alavani~\etal~\cite{alavani_program_2023} use static analysis of PTX instructions to train ML models for program-level post-execution energy estimation; however, their approach uses total thread counts and cannot capture block shape variations that cause significant energy differences~\cite{kerneltuner}, and requires extensive hardware-specific training data.
Cycle-accurate simulators such as 
\textit{GPUWattch}~\cite{leng_gpuwattch_2013} and \textit{AccelWatch}~\cite{kandiah2021accelwattch} provide microarchitectural details, but their computational overhead is prohibitive for exploring the vast config space of transformer kernels. 

\vspace{-1mm}
\paragraph{\textbf{GPU Kernel Optimization}}
\label{subsec:optimization}

Kernel optimization frameworks address energy-throughput trade-offs through config searches. 
For example, Kernel Tuner~\cite{schoonhoven_going_2022} uses runtime profiling to find optimal configs, while KLARAPTOR~\cite{brandt_klaraptor_2019} predicts parameters at runtime using analytical surrogates. 
Lou \& Muller~\cite{lou2024automatic} apply static analysis for CUDA kernel optimization but focus solely on performance without energy considerations, while Lim \etal{}~\cite{lim2017autotuning}
use simulation-based static analysis without power modeling or shape sensitivity.
Boughzala \etal{}’s approach~\cite{BoughzalaPredicting2020} focuses on thread block counts, ignoring shape effects and micro-architectural bottlenecks such as 
memory divergence, bank conflicts, and synchronization delays, which critically affect energy efficiency in \gpu{} workloads.

\vspace{-1mm}

\paragraph{\textbf{GPU Execution Challenges}}
\label{subsec:gpu-arch}

Transformer workloads face three challenges. First, memory coalescing efficiency~\cite{nvidiacudaguide2025} degrades due to non-contiguous access patterns in attention computations, increasing transaction costs. Second, maximizing SM occupancy through thread parallelism often conflicts with energy efficiency due to resource contention and sublinear power scaling~\cite{seyyedaghaei2024gpu}. Third, variable memory efficiency in attention kernels requires hybrid models to account for partial coalescing and warp scheduling dynamics, complicating traditional heuristic-based optimizations.

\vspace{-1mm}
\paragraph{\textbf{Comparison with existing work}}
Unlike existing approaches that rely on exhaustive runtime profiling or simplified models, \textit{FlipFlop leverages static PTX analysis to predict energy-efficient thread block configs without kernel execution}, reducing profiling overhead. It uniquely integrates thread block shape optimization with hardware-calibrated power modeling, addressing memory coalescing and power scaling challenges. This enables software developers to optimize kernel programs during design, offering explainable guidance that enhances energy efficiency across diverse GPU architectures. 
Table~\ref{tab:related-work-comparison} systematically compares FlipFlop with prior GPU energy optimization approaches across multiple dimensions: granularity, execution stage (pre-execution, runtime, or post-execution analysis), energy awareness, block shape sensitivity, explicit power modeling, pre-execution availability, and profiling requirements. We organize existing work into five categories based on their primary methodology: static analysis for GPUs, static analysis in other domains, analytical and simulation models, framework and runtime systems, and profiling-based autotuners. FlipFlop is distinguished by its unique combination of kernel-level granularity with pre-execution static analysis that simultaneously addresses energy optimization, block shape effects, and power modeling without requiring exhaustive profiling---capabilities not collectively present in any single prior approach. Notably, compilation frameworks like PyTorch Inductor~\cite{torchinductor}
and TensorRT~\cite{tensorRT}
operate at graph-level optimization and are complementary to our kernel-level approach: they optimize model structure while FlipFlop optimizes the launch configs of kernels that these frameworks execute.

\begin{table}[!ht]
\centering
\vspace{-2mm}
\caption{FlipFlop VS Prior GPU Energy Opt. Approaches. E=Energy-aware, S=Shape sensitivity, Pw=Power modeling, Pr=Pre-execution, Pf=Profile-free. 
}
\vspace{-4mm}
\label{tab:related-work-comparison}
\resizebox{0.75\columnwidth}{!}{
\small
\setlength{\tabcolsep}{3.5pt}
\renewcommand{\arraystretch}{1.0}
\rowcolors{2}{gray!8}{white}
\begin{tabular}{@{}llcccccc@{}}
\toprule
\rowcolor{white}
\textbf{Approach} & \textbf{Granularity} & \textbf{E} & \textbf{S} & \textbf{Pw} & \textbf{Pr} & \textbf{Pf} & \textbf{Type} \\
\midrule
\rowcolor{yellow!12}
\textbf{FlipFlop} & \textbf{Kernel} & \textbf{\cmark} & \textbf{\cmark} & \textbf{\cmark} & \textbf{\cmark} & \textbf{\cmark} & \textbf{Static} \\
\midrule
\rowcolor{gray!15}
\multicolumn{8}{l}{\textit{\textbf{Static Analysis for GPUs}}} \\
\midrule
Alavani et al.\cite{alavani_prediction_2023} & Program & \cmark & \xmark & \xmark & \xmark & \xmark & Static+ML \\
Lou \& Muller\cite{lou2024automatic} & Kernel & \xmark & \xmark & \xmark & \cmark & \cmark & Static \\
Lim et al.\cite{lim2017autotuning} & Kernel & \xmark & \xmark & \xmark & \xmark & \xmark & Static+Sim \\
\midrule
\rowcolor{gray!15}
\multicolumn{8}{l}{\textit{\textbf{Other Domain Static}}} \\
\midrule
Marantos et al.\cite{marantos2021flexible} & Program & \cmark & \xmark & \xmark & \cmark & \cmark & Static \\
Bangash et al.\cite{bangash2023energy} & Program & \cmark & \xmark & \xmark & \xmark & \cmark & Static \\
Grech et al.\cite{grech2015static} & Program & \cmark & \xmark & \xmark & \cmark & \cmark & Static \\
\midrule
\rowcolor{gray!15}
\multicolumn{8}{l}{\textit{\textbf{Analytical \& Simulation}}} \\
\midrule
Hong \& Kim\cite{hong2009analytical} & Kernel & \cmark & \xmark & \cmark & \cmark & \cmark & Analytical \\
GPUWattch\cite{leng_gpuwattch_2013} & Kernel & \cmark & \xmark & \cmark & \xmark & \xmark & Simulator \\
AccelWattch\cite{kandiah2021accelwattch} & Kernel & \cmark & \xmark & \cmark & \xmark & \xmark & Simulator \\
Boughzala et al.\cite{BoughzalaPredicting2020} & Kernel & \cmark & \xmark & \xmark & \xmark & \cmark & Static+Sim \\
\midrule
\rowcolor{gray!15}
\multicolumn{8}{l}{\textit{\textbf{Frameworks \& Runtime}}} \\
\midrule
PyTorch Inductor~\cite{torchinductor} & Framework & \xmark & \xmark & \xmark & \xmark & \xmark & Dynamic \\
TensorRT~\cite{tensorRT} & Framework & \xmark & \xmark & \xmark & \xmark & \xmark & Dynamic \\
NVIDIA Occ. Cal.\cite{nvidiaNsightCompute} & Kernel & \xmark & \xmark & \xmark & \cmark & \cmark & Heuristic \\
Zeus\cite{you2023zeus} & Framework & \cmark & \xmark & \cmark & \xmark & \xmark & Dynamic \\
Zamani et al.\cite{zamani_saou_2020} & System & \cmark & \xmark & \cmark & \xmark & \xmark & Dynamic \\
\midrule
\rowcolor{gray!15}
\multicolumn{8}{l}{\textit{\textbf{Profiling Autotuners}}} \\
\midrule
Kernel Tuner\cite{kerneltuner} & Kernel & \xmark & \cmark & \xmark & \xmark & \xmark & Profiling \\
KLARAPTOR\cite{brandt_klaraptor_2019} & Kernel & \xmark & \cmark & \xmark & \xmark & \xmark & Analytical \\
CLTune\cite{nugteren2015cltune} & Kernel & \xmark & \cmark & \xmark & \xmark & \xmark & Profiling \\
\bottomrule
\end{tabular}
}
\vspace{-3mm}
\end{table}

\vspace{-3mm}

\section{Overview}

This study aims to: (1) determine energy-efficient GPU configurations and power limits that maximize throughput; (2) predict these optima via static analysis; and (3) quantify savings in developer effort and environmental impact. Through three research questions and a unified experiment, we offer critical guidance for engineering sustainable AI systems.

\vspace{-2mm}
\subsection{Research Questions}

\textbf{RQ1: }\emph{How does tuning \gpu{} kernel thread block configurations and power limits dynamically affect energy efficiency and throughput of \llm{} inference, and what implications do these effects have for SE practices in AI system development?}
Software developers typically rely on default or heuristic-driven thread block configurations when writing CUDA kernels due to limited hardware expertise and abstraction barriers imposed by existing frameworks. In AI systems, especially LLMs, this challenge becomes acute due to highly variable sequence lengths during inference, each demanding specific kernel launch configurations for optimal performance. Multi-head attention (MHA) kernels represent a critical optimization target as they dominate computational costs in transformer architectures. Additionally, GPU power limits, which directly influence energy consumption, are usually set to vendor-provided defaults. Without proper tooling and awareness, software engineering teams inadvertently accept these defaults, potentially leading to substantial energy wastage. \textbf{RQ1} investigates how jointly adjusting thread block dimensions and GPU power limits optimizes energy efficiency and throughput for compute-intensive MHA kernels.

\vspace{1mm}
\noindent
\textbf{RQ2: }\emph{{Can static analysis predict optimal configurations that match dynamically-tuned energy optima?}}
Current optimization approaches require resource-intensive runtime profiling, creating barriers to energy-aware development. \textbf{RQ2} evaluates whether static analysis of PTX intermediate representations can accurately predict thread block configurations achieving energy efficiency comparable to dynamic tuning. This capability would enable developers to identify near-optimal configurations during design rather than through post-implementation profiling.

\vspace{1mm}
\noindent
\textbf{RQ3:} \emph{{How much developer effort and computational resources does static shape optimization save compared to profiling-based approaches?}}
Kernel optimization through exhaustive profiling creates substantial practical barriers: consuming significant time (hours to days), wasting computational resources, and increasing the carbon footprint. Provided static optimizations reduce adoption barriers (explored in RQ2), \textbf{RQ3} quantifies the savings in compute and time.

\vspace{-2mm}
\subsection{Experimental Setup}
\label{subsec:setup}

Our experiments used NVIDIA RTX 5000 Ada (24GB) and RTX 3070 Ampere (8GB) GPUs with AMD EPYC 9554P 64-core CPU, 1TB DDR4-3200 RAM, and 512MB L3 cache. Power was monitored via NVIDIA Management Library (NVML)~\cite{nvidiaNVIDIAManagement} on Ubuntu 20.04 with CUDA 12.4, Python 3.8, and modified \texttt{llama2.c}~\cite{githubGitHubKarpathyllama2c} framework for CodeLLAMA~\cite{llama2.cuda}. Profiling employed Nsight Compute~\cite{nvidiaNsightCompute}, Kernel Tuner~\cite{kerneltuner} with custom energy observers~\cite{kerneltunerObserversx2014}, and \texttt{pycuda} for GPU queries. SM utilization was tracked via CUDA Performance Tools Interface (CUPTI)~\cite{nvidiaCUPTIx2014}. Each configuration ran ten independent trials for stability. GPU power limits were adjusted in 25W increments from 100W to 250W (TDP).

\textbf{Kernels and Workloads:} We utilized diverse kernels representing critical AI computational patterns. General-purpose benchmarks from HeCBench~\cite{githubGitHubZjinlcfHeCBench}---vector addition, matrix multiplication, convolution, reduction, scalar product, and transpose---serve as fundamental building blocks for AI workloads. LLM kernels focusing on multi-head attention (MHA) operations from LLAMA~\cite{llama3.cuda} exhibit quadratic complexity ($O(L^2)$ for sequence length $L$), making them key targets for energy optimization. Specialized kernels, such as 3D Laplace solvers and custom attention kernels, address scientific computing and emerging AI workloads.
For each experiment, we generated valid thread block configurations by enumerating combinations of \texttt{block\_size\_x} and \texttt{block\_size\_y} (14 values: 1, 2, 4, ..., 1024) and power limits (10 values: 100W to 250W). We filtered these to ensure total threads per block (\texttt{block\_size\_x} $\times$ \texttt{block\_size\_y}) ranges from 32 to 1024 and is divisible by 32, adhering to NVIDIA GPU warp-alignment constraints. This yields a subset of valid configurations per sequence length (\eg{} 121 for RQ1, 66 for RQ3), with example factorizations such as [1$\times$32], [2$\times$16] for 32 threads and [1$\times$1024], [1024$\times$1] for 1024 threads, enabling rigorous evaluation of geometry effects and efficient optimization.


\vspace{-1mm}
\section{RQ1: Tuning Thread Blocks and Power Limits}
\label{sec:rq1}

\noindent
\paragraph{\textbf{Methods---RQ1}}

\vspace{-2mm}
We evaluate MHA kernels across sequence lengths $\{128, 256, \dots 8192\}$. We test all valid thread block configs with x- and y-dimensions, each drawn from $\{1, 2, 4, \dots 1024\}$, constrained so that the total number of threads ranges from 32 to 1024 and is divisible by 32, yielding 121 configs. Additionally, we assess seven power limits (100W-250W, at 25W increments), adjusted dynamically using NVML~\cite{nvidiaNVIDIAManagement}, to explore energy-performance trade-offs in realistic LLM inference settings.

To emulate real-world LLM inference, we create a partial-decoding loop where the MHA kernel is launched repeatedly as tokens are generated. For each decoding step, we exhaustively evaluate all valid block configs and power limit combinations to identify the most energy-efficient setup. We measure energy usage, execution time, and throughput (tokens/s) over ten repeated runs for reliability. This process iterates across multiple decoding steps to improve measurement accuracy.

\vspace{-2mm}
\paragraph{\textit{Data collection and metrics.}} 
We evaluate energy efficiency in terms of \emph{Joules per token} (J/token) and assess computational performance through throughput (tokens/s) and latency per token measurements.
During each kernel launch, we track GPU instantaneous power through NVML sampling at 1ms intervals. We integrate these measurements over kernel execution duration to compute total energy consumption, then normalize by processed tokens to obtain \emph{Joules per token}: \emph{Joules per token}: $\text{J/tok} = \frac{\sum P(t_i) \cdot \Delta t}{\text{batch} \times \text{L} \times N_{\text{runs}}}$,
where $P(t_i)$ is instantaneous power and $N_{\text{runs}}=10$ ensures measurement stability. CUPTI validates SM occupancy and warp scheduling efficiency through the \texttt{sm\_warps\_active} metric.
Additionally, we measure SM efficiency as the ratio of average active warps to the GPU's maximum supported active warps, providing insights into hardware utilization beyond traditional occupancy metrics. To contextualize findings, we benchmark results against static heuristic configs from NVIDIA's Occupancy Calculator operating at default power limits.

\vspace{-1mm}
\noindent
\paragraph{\textbf{Results---RQ1}}

\begin{figure}[t]
    \centering
    \vspace{-4mm}
    \includegraphics[width=\columnwidth]{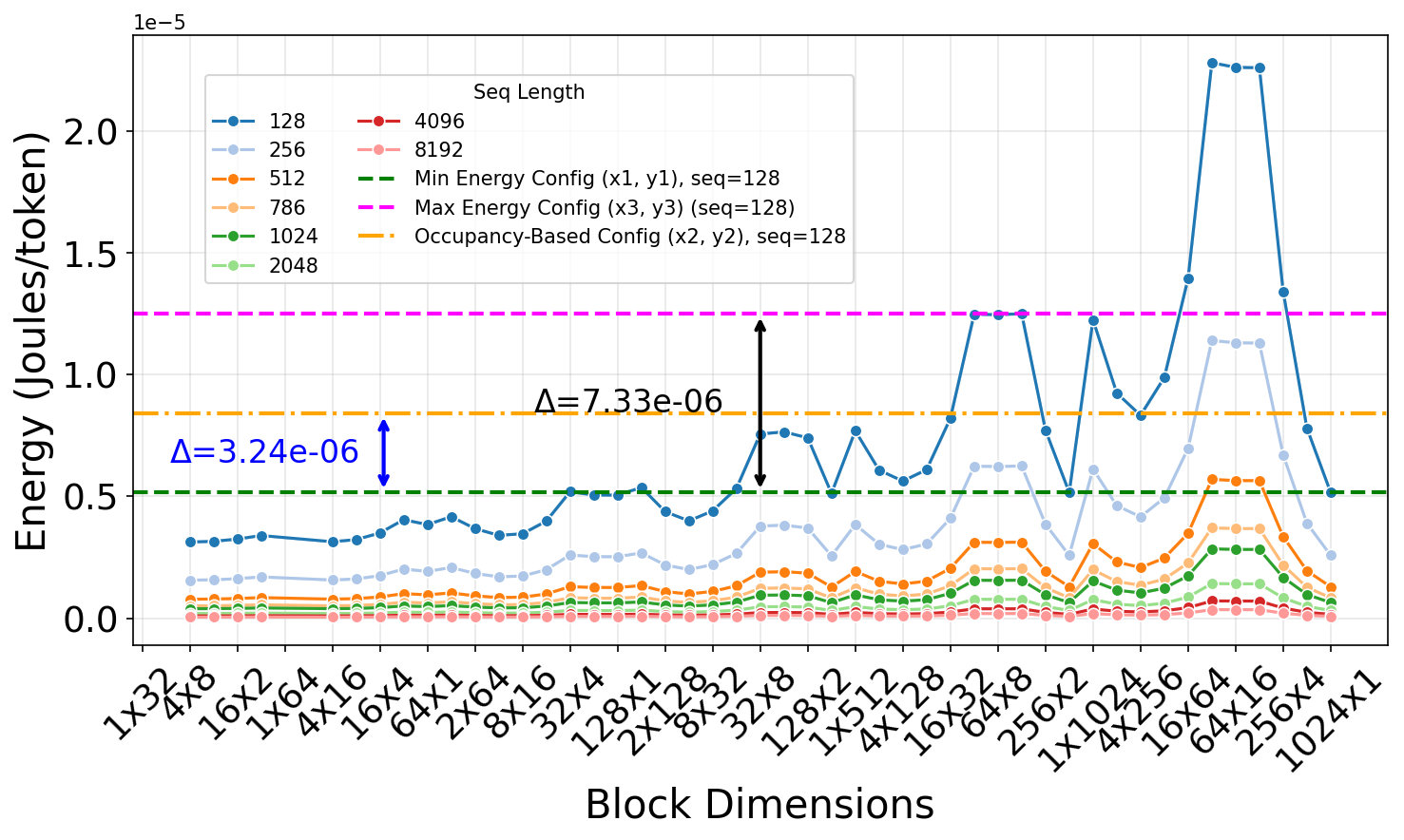}
    \vspace{-2em}
    \caption{Energy per token vs. block dimensions.  Each line represents a sequence length. The horizontal line shows the consumed energy of occupancy-based configs}
    \label{fig:rq1-energy-blockdims}
    \vspace{-5mm}
\end{figure}

Figure~\ref{fig:rq1-energy-blockdims} displays energy consumed per token (J/token) as a function of block config. A horizontal line marks baseline energy from NVIDIA's Occupancy Calculator at default power (250W). The x-axis is organized by total threads per block in increasing order, then by increasing thread block x-dimension (\eg{} for 32 threads: (1$\times$32), (2$\times$16), (4$\times$8), etc.). Different sequence lengths are shown using different line colors to capture how workload size impacts energy efficiency.

\vspace{-1mm}
\vspace{-1mm}

\paragraph{Comparison with Static Heuristics.} Figure~\ref{fig:rq1-energy-blockdims} depicts the occupancy-based heuristic for reference. At sequence length 128, the Occupancy Calculator recommends a 512-thread config (\eg{} \texttt{16x32}), yielding 100\% theoretical occupancy. However, our adaptive search finds a shape using about $3.24\mu J$ \emph{less} energy per token compared to that occupancy-based choice and can save up to $7.33\mu J$ relative to the \emph{least} efficient shape tested. \textbf{Hence, while occupancy maximization may be computationally appealing, it does not reliably provide the lowest energy per token in practice.}

\vspace{-2mm}
\paragraph{Per-Sequence Length Analysis.}
Table~\ref{tab:rq2-best-configs} provides detailed insights into energy efficiency across sequence lengths from 128 to 8,192, showcasing exhaustive tuning impacts on block configs and power limits. Critically, these results demonstrate that optimal configs fundamentally depend on both input size (sequence length) and thread block shape---not just total thread count. At sequence length 512, the optimal config at 100W with 2×32 block shape achieves $0.88\times10^{-6}$ J/token, a 78.9\% reduction from the 250W baseline of $3.87\times10^{-6}$ J/token, while maintaining throughput of $1.24\times10^{5}$ tokens/s. Similarly, at 256, the best config at 100W with 2×32 shape yields $1.63\times10^{-6}$ J/token, saving $3.39\times10^{-6}$ J/token (67.5\% reduction) with throughput boost to $0.62\times10^{5}$ tokens/s. Lower power limits (100–165W) consistently outperform the 250W baseline, with energy savings ranging from 8.3\% at 128 to 78.9\% at 512, demonstrating the efficacy of shape-aware optimization over static thread count maximization or occupancy-based heuristics. Although absolute energy savings vary with sequence length, the trend is clear: jointly adapting GPU power cap and block shape realizes substantial energy savings without compromising throughput. \textbf{These findings indicate that accurately predicting energy-efficient configs requires modeling the interplay between input size, block shape dimensions, and memory access patterns}---parameters that collectively determine kernel behavior under varying workload conditions.

\begin{table}[ht!]
\centering
\vspace{-2mm}
\caption{Optimal configs per Sequence Length vs. Baseline. 
$\Delta E = (E_{\text{base}} - E_{\text{opt}})/E_{\text{base}} \times 100\%$; $\Delta \text{Thr} = (\text{Thr}_{\text{opt}} - \text{Thr}_{\text{base}})/\text{Thr}_{\text{base}} \times 100\%$. Rounded mean of 10 runs($\sigma<5\%$).}
\vspace{-2mm}

\resizebox{0.9\columnwidth}{!}{
\label{tab:rq2-best-configs}
\footnotesize
\setlength{\tabcolsep}{2.5pt} 
\begin{tabular}{cccccccc}
\toprule
\textbf{Seq.} & \textbf{Power} & \textbf{Block} & \textbf{J/tok} & \textbf{Thr.} & \textbf{Base.} & \textbf{$\Delta$ E} & \textbf{$\Delta$ Thr.} \\
\textbf{Len} & \textbf{(W)} & \textbf{(x$\times$y)} & \textbf{(J/tok)} & \textbf{(tok/s)} & \textbf{(J/tok)} & \textbf{(\%)} & \textbf{($\pm$\%)} \\
\midrule
128  & 135 & 8$\times$8  & $3.10\times10^{-6}$ & $0.33\times10^{5}$ & $3.38\times10^{-6}$ & 8.3  & -1.3 \\
256  & 100 & 2$\times$32 & $1.63\times10^{-6}$ & $0.62\times10^{5}$ & $5.03\times10^{-6}$ & 67.5 & +143.2 \\
512  & 100 & 2$\times$32 & $0.88\times10^{-6}$ & $1.24\times10^{5}$ & $3.87\times10^{-6}$ & 78.9 & +124.1 \\
786  & 135 & 8$\times$8  & $0.51\times10^{-6}$ & $2.02\times10^{5}$ & $0.69\times10^{-6}$ & 26.7 & +35.0 \\
1024 & 100 & 2$\times$32 & $0.41\times10^{-6}$ & $2.48\times10^{5}$ & $1.25\times10^{-6}$ & 67.4 & +175.3 \\
2048 & 165 & 4$\times$16 & $0.12\times10^{-6}$ & $5.49\times10^{5}$ & $0.25\times10^{-6}$ & 22.3 & +9.4 \\
4096 & 100 & 2$\times$32 & $0.12\times10^{-6}$ & $9.90\times10^{5}$ & $0.40\times10^{-6}$ & 74.7 & +302.0 \\
8192 & 100 & 2$\times$32 & $0.06\times10^{-6}$ & $19.8\times10^{5}$ & $0.08\times10^{-6}$ & 38.5 & +58.5 \\
\bottomrule
\end{tabular}
}
\vspace{-3mm}
\end{table}
\vspace{-2mm}

\paragraph{\textbf{Understanding the Energy-Runtime-Power Relationship}}
\label{sec:rq1-energy-runtime}

Having established that joint optimization yields substantial energy savings (Table~\ref{tab:rq2-best-configs}), we investigate the underlying relationship between energy consumption, execution time, and power draw. To rigorously examine this relationship, we conduct an expanded evaluation across 464 unique thread block configs (ranging from $1\times1$ to $1024\times1024$ threads) tested at 8 sequence lengths without power limits, producing 11,040 measurements. This broader exploration examines whether runtime alone is sufficient to predict energy, or if additional factors need to be considered.
We examine energy-runtime correlation across three progressively broader config subsets.
\textbf{Our correlation analysis reveals a critical insight.}
First, within optimal configs (Table~\ref{tab:rq2-best-configs}), we observe strong correlation ($R^2 = 0.91$, $p < 0.001$), where 91\% of energy variance correlates with execution time. Second, examining 1,601 pairwise comparisons between baseline and candidate configs reveals substantially weaker correlation ($r = 0.46$, $R^2 = 0.21$, $p < 0.001$), with 78.8\% unexplained variance---runtime explains less than one-quarter of energy variation. Third, across the \emph{complete} config space of all 11,040 measurements, correlation becomes \emph{negative} ($r = -0.17$, $p < 0.001$), where poor configs exhibit both high runtime \emph{and} high power simultaneously. This three-level analysis demonstrates that while optimal configs exhibit predictable energy-time relationships, \emph{discovering} these configs from the exhaustive space requires explicit consideration of factors beyond runtime.
\textbf{Power draw variability} explains why runtime alone fails. Power draw exhibits substantial variability: up to 5.11$\times$ across sequence lengths and 2.07$\times$ among thread block configs. This variation stems from three architectural mechanisms: coalesced memory accesses minimize power consumption by enhancing memory efficiency~\cite{crago2018exposing,nvidiacudabestpractice2025},
warp divergence and resource contention introduce power fluctuations uncorrelated with throughput due to execution unit idleness~\cite{sadrosadati2019itap,kandiah2021accelwattch}, and dynamic voltage-frequency scaling induces power shifts as the GPU adjusts to workload intensities~\cite{guerreiro2019modeling}. These mechanisms produce complex power-config interactions invisible to runtime-only approaches.
\textbf{Greenup-Speedup analysis}~\cite{abdulsalam2015using} quantifies how energy efficiency diverges from performance. Analysis across 1,601 config comparisons reveals statistically significant divergence: mean 9.64\% (range 0\%--51.70\%), with 39.7\% of cases exceeding 5\% divergence ($p < 0.001$) and 31.2\% exceeding 10\%. Critically, 18.0\% of cases exhibit 20--50\% divergence, with extreme cases reaching 51.70\% where configs achieve 3.85$\times$ speedup but 2.07$\times$ powerup, yielding only 1.86$\times$ greenup. This demonstrates that power varies independently of runtime, creating energy-saving opportunities invisible to time-only optimization.
\textbf{Implications for optimization strategy.} The statistically significant degradation from strong correlation in optimal configs ($R^2 = 0.91$) to moderate in pairwise comparisons ($R^2 = 0.21$) to negative in complete space ($r = -0.17$, all $p < 0.001$) demonstrates that energy-optimal configs cannot be reliably identified through runtime analysis alone. The 78.8\% unexplained variance quantifies the extent to which time-only approaches fail to capture critical power dynamics. Crucially, while runtime correlates strongly with energy for \emph{known} optimal configs (91\% variance explained), one cannot use runtime correlation to \emph{prospectively identify} which configs will be optimal from thousands of candidates, where 39.7\% of config pairs show substantial divergence between energy and performance improvements. Our joint tuning of power limits and thread block configs addresses this limitation, enabling navigation from the chaotic complete space ($r = -0.17$) to optimal configs. The prevalence of substantial divergence (18.0\% of cases exceeding 20\%), wide power variation (up to 2.07$\times$), and minimal variance explained by runtime alone (21\%) demonstrates that energy-aware GPU optimization must treat power and time as coupled but distinct objectives.

We summarize \textbf{key observations} from the results below.

\begin{itemize}[leftmargin=*]
    \item \textit{Non-monotonic trend:} Developers often expect higher thread block dimensions to reduce energy usage by maximizing SM occupancy, as occupancy-based studies suggest that increased parallelism hides memory latency and improves efficiency~\cite{volkov2010better}. However, our results show that best energy efficiency often occurs at moderate thread counts (\eg{} 128--384 threads per block), suggesting configs maximizing occupancy may not optimally exploit other resources such as shared memory or registers.
    
    \item \textit{Adaptive advantage:} Comparing our adaptive search with NVIDIA's Occupancy Calculator (Figure~\ref{fig:rq1-energy-blockdims}), the occupancy-based approach (typically recommending high thread counts, \eg{} 256 threads per block) does not lie on the Pareto front. Instead, adaptive search frequently discovers configs that both lower energy by 15--20\% per token(representing typical mid-range savings across sequence lengths when comparing our adaptive optima against occupancy baselines in Figure~\ref{fig:rq1-energy-blockdims}) and improve FLOPS/Watt.
    
    \item \textit{Throughput and utilization:} Despite lower SM occupancy at times, thread block configs found by our adaptive approach achieve higher overall energy efficiency, indicating that maximizing occupancy does not necessarily translate into optimal energy usage.
\end{itemize}

\vspace{-1mm}
\paragraph{\textit{Answer to RQ1.}}
\emph{Yes---jointly tuning power limits and thread block configs reveals energy-optimal operating points invisible to either parameter alone. Across sequence lengths 128--8,192, adaptive configs (typically 100--165W with geometry-specific block shapes) reduce energy per token by up to 79\% versus occupancy-based heuristics at default power, while maintaining >95\% throughput. Optimal configs depend fundamentally on both input size and thread block shape dimensions---not just total thread count---with counterintuitive patterns such as 2×32 at 100W achieving 78.9\% energy savings at seq\_len=512 versus occupancy's 512-thread recommendation. Statistical analysis (Section~\ref{sec:rq1-energy-runtime}) validates necessity: energy-runtime correlation degrades from $R^2 = 0.91$ in optimal configs to $R^2 = 0.21$ in pairwise comparisons to $r = -0.17$ across complete design space ($p < 0.001$), with 39.7\% of config pairs showing >5\% Greenup-Speedup divergence---demonstrating that runtime-only approaches systematically fail to identify energy-efficient configs. However, exhaustive profiling of such joint search spaces remains prohibitively expensive, highlighting the need for predictive approaches.}

\vspace{-1mm}
\section{RQ2: Predict and Optimize Power and Energy } \label{sec:RQ2}

\vspace{-1mm}

\noindent
\paragraph{\textbf{Methods---RQ2}}

To enable developers to create energy-efficient GPU applications without extensive runtime profiling, we developed the \emph{FlipFlop} framework. The framework predicts optimal kernel configs using static analysis of PTX representation of GPU kernel code, tackling the challenge of balancing energy consumption and performance in GPU-based AI software systems where power efficiency is critical. By analyzing PTX and combining it with hardware-specific calibration, \emph{FlipFlop} abstracts complex hardware details, allowing developers to optimize energy usage with minimal hardware expertise.

\begin{algorithm}[t]
\caption{FlipFlop Energy Prediction Model}
\label{alg:flipflop}
\scriptsize
\linespread{0.85}\selectfont
\SetAlgoLined
\DontPrintSemicolon
\SetAlCapSkip{0.5em}

\KwIn{PTX code, GPU architecture $\mathcal{A}$, configuration $\mathcal{C} = (\mathtt{block\_x}, \mathtt{block\_y}, P_{\mathtt{cap}})$,}
input resources $\mathcal{R}$ (shared\_mem, grid\_dims)\;
\KwOut{Predicted energy $E_{\mathtt{pred}}$}

\textbf{// Calibration Phase} $\{\beta_u, L_{\mathtt{mem}}^{\mathtt{coal}},...\}$\;

\textbf{// Phase 1: Static Feature Extraction}\;
$\mathcal{F} \gets \text{ExtractFeatures}(\text{PTX}, \mathcal{C}, \mathcal{R})$\;
$\mathcal{F}.\eta_{\mathtt{coal}} \gets \min\left(1, \frac{\mathcal{C}.\mathtt{block\_x}}{32}\right) \times \frac{\text{aligned accesses}}{\text{total accesses}}$ \tcp*[r]{Eq. (1)}
$\mathcal{F}.N_{\mathtt{mem}} \gets \text{count(load/store instructions)}$\;
$\mathcal{F}.N_{\mathtt{comp}} \gets \text{count(FP32/INT/SFU instructions)}$\;
$\mathcal{F}.N_{\mathtt{sync}} \gets \text{count}(\texttt{.sync} \text{ instructions})$\;
$\mathcal{F}.\mathtt{warps} \gets \frac{\mathcal{C}.\mathtt{block\_x} \times \mathcal{C}.\mathtt{block\_y}}{32}$\;
$\mathcal{F}.\mathtt{blocks\_per\_SM} \gets \min\left(\frac{\text{MAX\_WARPS}}{\mathcal{F}.\mathtt{warps}}, \frac{\text{MAX\_SHARED}}{\mathcal{R}.\text{shared\_mem}}\right)$ \tcp*[r]{Input-dependent occupancy}

\textbf{// Phase 2: Execution Time Estimation}\;
$T_{\mathtt{exec}} \gets 0$\;
$\mathtt{MWP} \gets L_{\mathtt{mem}}^{\mathtt{coal}} / \mathtt{departure\_delay}$ \tcp*[r]{Hong-Kim model~\cite{hong2009analytical}}
$\mathtt{CWP} \gets (\mathtt{cycles_{mem}} + \mathtt{cycles_{comp}}) / \mathtt{cycles_{comp}}$\;
$BW_{\mathtt{eff}} \gets BW_{\max} \times \mathcal{F}.\eta_{\mathtt{coal}}$\;
$T_{\mathtt{mem}} \gets \mathcal{F}.N_{\mathtt{mem}} / (\mathtt{MWP} \times BW_{\mathtt{eff}})$\;
$T_{\mathtt{comp}} \gets \mathcal{F}.N_{\mathtt{comp}} / (\mathtt{CWP} \times \mathtt{IPC})$\;
$T_{\mathtt{sync}} \gets \mathcal{F}.N_{\mathtt{sync}} \times t_{\mathtt{barrier}}$\;
$T_{\mathtt{exec}} \gets \alpha T_{\mathtt{mem}} + \beta T_{\mathtt{comp}} + \gamma T_{\mathtt{sync}} + T_{\mathtt{base}}$ \tcp*[r]{Eq. (2)}

\textbf{// Phase 3: Dynamic Power Estimation}\;
$P_{\mathtt{dyn}} \gets 0$\;
\ForEach{unit $u \in \{\mathtt{FP32}, \mathtt{INT}, \mathtt{SFU}, \mathtt{ALU}, \mathtt{Mem}\}$}{
    $\mathtt{AR}_u \gets \mathcal{F}.\mathtt{inst\_count}(u) \times \frac{\mathtt{warps\_per\_SM}}{\mathtt{exec\_cycles}/\mathtt{issue\_cycles}}$\;
    $P_{\mathtt{dyn}} \gets P_{\mathtt{dyn}} + \beta_u \times \mathtt{AR}_u$ \tcp*[r]{$\beta_u$ from calibration}
}
$\mathtt{CI} \gets \mathcal{F}.N_{\mathtt{comp}} / \mathcal{F}.N_{\mathtt{mem}}$ \tcp*[r]{Compute intensity}
$P_{\mathtt{shape}} \gets P_{\mathtt{base}} \times \left(1 + \frac{\kappa \left|\ln\left(\frac{\mathcal{C}.\mathtt{block\_x}}{\mathcal{C}.\mathtt{block\_y}}\right)\right|}{1 + \mathtt{CI}}\right)$ \tcp*[r]{Eq. (4)}
$P_{\mathtt{mem}} \gets P_{\mathtt{mem}}^{\mathtt{base}} \times \left(1 + \lambda(1 - \mathcal{F}.\eta_{\mathtt{coal}})\right)$ \tcp*[r]{Eq. (5)}
$P_{\mathtt{dyn}} \gets P_{\mathtt{dyn}} + P_{\mathtt{shape}} + P_{\mathtt{mem}}$\;
$n \gets \text{estimate\_active\_SMs}(\mathcal{C}, \mathcal{A})$\;
$P_{\mathtt{SM}} \gets \alpha n^{\beta} + \delta$ \tcp*[r]{Concurrency scaling}
$P_{\mathtt{dyn}} \gets P_{\mathtt{dyn}} + P_{\mathtt{SM}}$\;
\If{$T_{\mathtt{exec}} < \tau_{\mathtt{short}}$}{
    $P_{\mathtt{dyn}} \gets P_{\mathtt{dyn}} \times r$ \tcp*[r]{Transient correction}
}

\textbf{// Phase 4: Energy Prediction}\;
$f_{\mathtt{adj}} \gets f_{\mathtt{base}} \times \left(\frac{\mathcal{C}.P_{\mathtt{cap}}}{P_{\mathtt{TDP}}}\right)^{1/3}$ \tcp*[r]{DVFS adjustment}
$P_{\mathtt{dyn}} \gets P_{\mathtt{dyn}} \times \frac{f_{\mathtt{adj}}}{f_{\mathtt{base}}}$\;
$E_{\mathtt{pred}} \gets T_{\mathtt{exec}} \times (P_{\mathtt{dyn}} + P_{\mathtt{static}}) + E_{\mathtt{overhead}}$\;
\KwRet $E_{\mathtt{pred}}$\;
\end{algorithm}

\begin{algorithm}[t]
\caption{Configuration Space Exploration}
\label{alg:explore}
\scriptsize
\linespread{0.85}\selectfont
\SetAlgoLined
\DontPrintSemicolon
\SetAlCapSkip{0.5em}

\KwIn{PTX code, GPU architecture $\mathcal{A}$, input dimensions $\mathcal{I}$ (seq\_len, batch, heads)}
\KwOut{Pareto-optimal configurations $\mathcal{P}$}

$\mathcal{P} \gets \emptyset$\;
$\mathcal{R} \gets \text{ComputeInputResources}(\mathcal{I})$ \tcp*[r]{shared\_mem, grid\_dims from input}
$\mathcal{S} \gets \text{GenerateValidConfigs}(\mathcal{A}, \mathcal{R})$ \tcp*[r]{HW + input constraints}

\ForEach{configuration $\mathcal{C} \in \mathcal{S}$}{
    $E_{\mathtt{pred}} \gets \text{PredictEnergy}(\text{PTX}, \mathcal{A}, \mathcal{C})$\;
    $T_{\mathtt{pred}} \gets \text{EstimateExecutionTime}(\cdots)$ \tcp*[r]{From Phase 2}
    \If{$\nexists \mathcal{C}' \in \mathcal{P} \text{ s.t. } E_{\mathtt{pred}}' \leq E_{\mathtt{pred}} \land T_{\mathtt{pred}}' \geq \rho T_{\mathtt{peak}}$}{
        $\mathcal{P} \gets \mathcal{P} \cup \{(\mathcal{C}, E_{\mathtt{pred}}, T_{\mathtt{pred}})\}$\;
    }
}
\KwRet $\mathcal{P}$\;
\end{algorithm}

\begin{algorithm}[t]
\caption{Architecture Calibration}
\label{alg:calibration}
\scriptsize
\linespread{0.85}\selectfont
\SetAlgoLined
\DontPrintSemicolon
\SetAlCapSkip{0.5em}

\KwIn{GPU $\mathcal{A}$}
\KwOut{Calibrated parameters}

\textbf{// Power characterization via microbenchmarks}\;
\ForEach{unit $u \in \{\mathtt{FP32}, \mathtt{INT}, \mathtt{Mem}, ...\}$}{
    $\beta_u \gets \frac{P_{\text{saturated}} - P_{\text{idle}}}{\text{max operations/s}}$ \tcp*[r]{Unit power coefficient}
}

\textbf{// Memory profiling}\;
$L_{\mathtt{mem}}^{\mathtt{coal}} \gets \text{measure\_latency}(\text{stride=1})$ \tcp*[r]{Coalesced}
$L_{\mathtt{mem}}^{\mathtt{uncoal}} \gets \text{measure\_latency}(\text{stride=128})$ \tcp*[r]{Uncoalesced}

\textbf{// Concurrency scaling}\;
\For{$n \gets 1$ \KwTo $\text{max SMs}$}{
    $P_{\text{SM}}(n) \gets \text{measure\_power}(\text{grid\_size}=n)$\;
    $\alpha_c, \beta_c, \delta_c \gets \text{fit}(P_{\text{SM}} = \alpha_c n^{\beta_c} + \delta_c)$\;
}

\textbf{// Transient effects}\;
$P_{\text{sustained}} \gets \text{measure}(n=100)$\;
$P_{\text{short}} \gets \text{measure}(\text{single kernel})$\;
$r \gets P_{\text{short}} / P_{\text{sustained}}$\;

\textbf{// Shape/coalescing coefficients}\;
$\kappa, \lambda \gets \text{fit via aspect ratio \& stride sweeps}$\;

\KwRet $\{\beta_u, L_{\mathtt{mem}}^{\mathtt{coal}}, ...\}$\;
\end{algorithm}

\begin{figure}[t!]
    \centering
    \vspace{-2mm}
    \includegraphics[width=0.9\columnwidth]{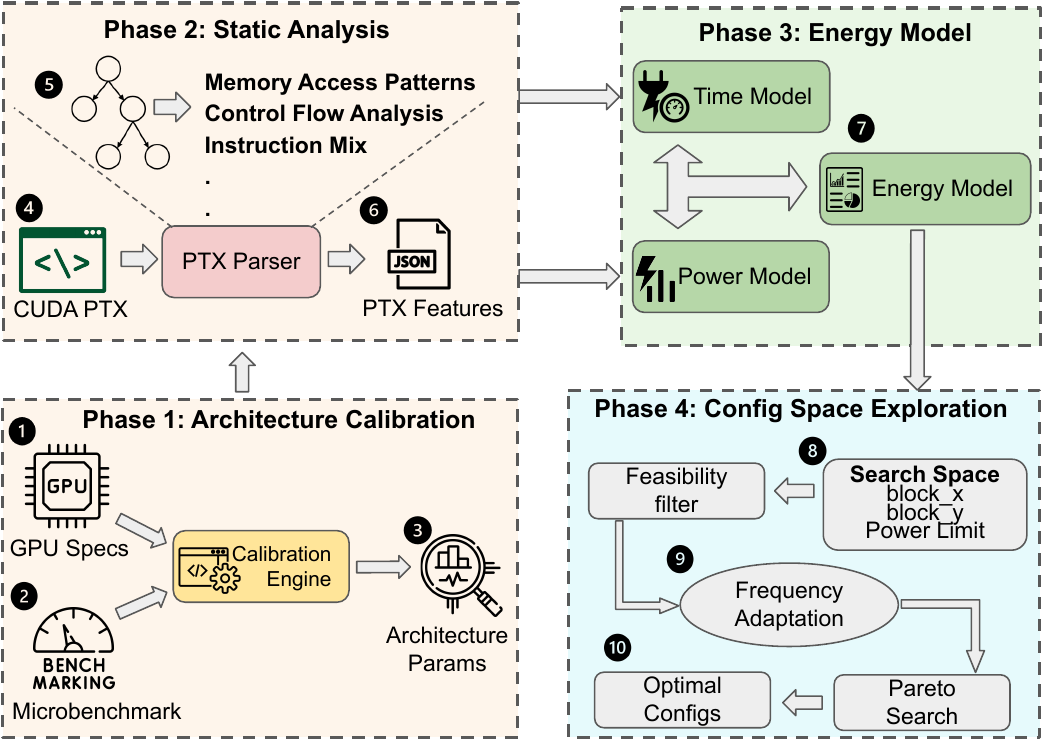}
    \vspace{-4mm}
    \caption{\emph{FlipFlop} Architecture}
    \label{fig:workflow}
    \vspace{-2mm}
\end{figure}

The \emph{FlipFlop} framework employs a four-phase pipeline (Figure~\ref{fig:workflow}) to predict energy-efficient GPU kernel configs, addressing three key challenges: predicting thread block geometry and power limit interactions during design, maintaining accuracy without costly runtime measurements, and generalizing optimizations across diverse GPU architectures. 
We elaborate on the pipeline and its key components in the rest of the section.

The framework integrates input size throughout the pipeline via \texttt{ComputeInputResources} (Algorithm~\ref{alg:explore}, line 2), which computes kernel-specific resource requirements that directly constrain config validity. Shared memory requirements scale with input dimensions---for instance, our MHA kernel implementation requires shared memory proportional to both the head dimension 
and sequence length---directly limiting blocks-per-SM as inputs grow. Grid dimensions similarly scale with workload geometry, determining total compute demand. The \texttt{GenerateValidConfigs} function (line 3) filters configs against both hardware limits (registers, warp count) and these input-dependent constraints, as configs valid for short sequences may violate shared memory limits for longer inputs. This input-aware filtering propagates to execution time modeling (Algorithm~\ref{alg:flipflop}, lines 9, 14), where occupancy modulates warp-level parallelism, capturing input size effects through hardware resource modeling.

\vspace{-1mm}

\paragraph{\textit{Architecture-specific calibration}}
\label{subsec:calibration}

\begin{figure}[t!]
    \centering
    \vspace{-4mm}
    \includegraphics[width=0.65\columnwidth]{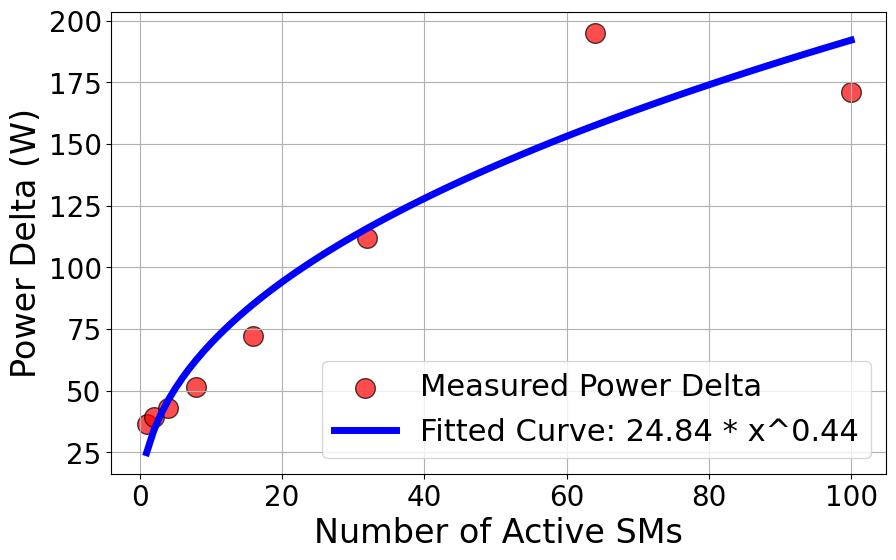}
    \vspace{-4mm}
    \caption{Power scaling vs. active SMs on RTX 5000 Ada.}
    \label{fig:concurrency-scaling}
    \vspace{-4mm}
    \end{figure}

\textbf{To ensure accurate energy predictions across diverse \gpu{} architectures, FlipFlop calibrates hardware-specific parameters, such as memory access latencies, concurrency limits, and per-instruction power consumption, through a systematic process.} This calibration abstracts low-level hardware complexities, allowing developers to reason about energy impacts without deep \gpu{} expertise. The process involves three stages. First, power characterization dynamically adjusts \gpu{} power limits (\eg{} 100W to 250W on an RTX 5000 \gpu{}) and uses targeted microbenchmarks to isolate energy contributions from functional units, such as floating-point (FP32) cores, integer ALUs, and memory controllers. This yields power coefficients (\(\beta_u\), Algorithm~\ref{alg:calibration}, line 3), enabling the model to adapt to hardware-specific behaviors without runtime measurements. Second, memory subsystem profiling measures latencies for coalesced (efficient) and uncoalesced (inefficient) memory accesses (\(L_{\text{mem}}^{\text{coal}}\), \(L_{\text{mem}}^{\text{uncoal}}\)), quantifies shared memory bank conflicts~\cite{nvidiaUsingShared}, and profiles effective bandwidth via sustained VRAM copy operations. This translates memory access patterns into energy costs, guiding developers to optimize memory usage. Third, concurrency scaling models power consumption as a function of active streaming multiprocessors (SMs) using a power-law relationship (\(P_{\text{SM}}(n) = \alpha n^\beta + \delta\))~\cite{seyyedaghaei2024gpu}, capturing diminishing returns due to resource contention and dynamic voltage and frequency scaling (DVFS) limitations (as shown in Figure~\ref{fig:concurrency-scaling}). Additionally, transient power effects for short-duration kernels (\eg{} 10\(\mu\)s) are quantified with a scaling factor (\(r = P_{\text{short}} / P_{\text{sustained}} = 0.833\)), preventing overestimation of power for bursty workloads (Algorithm~\ref{alg:flipflop}, line 32). Calibration ensures a portable and accurate energy model, enabling developers to focus on writing efficient code.

    \vspace{-1mm}
    
\paragraph{\textit{Static kernel feature extraction}}
\label{subsec:static-features}

To predict kernel behavior before execution, \emph{FlipFlop} analyzes PTX code.
We choose PTX over SASS (hardware-specific assembly) because: (1) PTX enables architecture-agnostic analysis across NVIDIA GPU generations without retraining, whereas SASS requires per-architecture models; and (2) our calibration phase (Section~\ref{subsec:calibration}) bridges PTX's abstraction gap by directly measuring hardware-specific behaviors (memory latencies, coalescing efficiency, power scaling) on target GPUs, capturing low-level details while maintaining portability.

The analysis proceeds in three stages. First, control flow reconstruction builds a control flow graph (CFG)~\cite{cfg2025}, mapping execution paths by identifying basic blocks, branches, and loops to estimate dynamic instruction counts crucial for understanding performance and computational intensity. Second, memory access characterization classifies memory operations as coalesced or uncoalesced based on access patterns~\cite{nvidiacudaguide2025}, computing a coalescing efficiency metric (\(\eta_{\text{coal}}\), Algorithm~\ref{alg:flipflop}, line 4) that quantifies how well memory accesses merge into fewer transactions, reducing overheads. Shared and local memory usage is analyzed via instruction pattern matching~\cite{nvidiaPtx2025}. Third, computational profile estimation categorizes instructions into floating-point (FMA), integer, special function units (SFU), and ALU operations, identifying compute-intensive regions and resource demands. Loop iteration counts and resource demands (\eg{} registers, shared memory) are estimated from compiler outputs. \emph{FlipFlop} determines whether a kernel is memory-bound or compute-bound by integrating these static metrics with runtime-derived Memory Warp Parallelism (MWP) and Compute Warp Parallelism (CWP)~\cite{hong_integrated_2010, hong2009analytical}.

\vspace{-1mm}
\paragraph{\textit{Performance-Time Modeling}}
\label{subsec:time-model}

The performance-time model predicts kernel execution time by analyzing the interplay between memory and compute operations, helping developers understand how config choices, such as thread block size and power cap, impact runtime. Building on the MWP-CWP framework~\cite{hong_integrated_2010, hong2009analytical}, the model accounts for modern GPU features such as 
partial memory coalescing and concurrency constraints. It integrates static code features extracted from PTX code, a low-level GPU instruction representation (Section~\ref{subsec:static-features}), with hardware-specific calibration data (Section~\ref{subsec:calibration}) to compute execution time (\(T_{\text{exec}}\)) for specific configs.
Implemented in Algorithm~\ref{alg:flipflop}, lines 14–17, the model calculates \(T_{\text{exec}}\) as a weighted sum of four components: memory access time (\(T_{\text{mem}}\)), compute time (\(T_{\text{comp}}\)), synchronization overhead (\(T_{\text{sync}}\)), and fixed kernel launch overhead (\(T_{\text{base}}\)), with weights (\(\alpha\), \(\beta\), \(\gamma\)) derived from calibration to reflect kernel characteristics. Memory subsystem modeling extracts memory operations, such as load and store instructions, from PTX code~\cite{nvidiaPtx2025}, estimating access times based on coalescing efficiency (\(\eta_{\text{coal}}\)) and VRAM bandwidth, and incorporates calibrated latencies for global, shared, and local memory. Compute pipeline modeling evaluates instruction throughput for floating-point (\eg{} FMA), integer, special function unit (SFU), and ALU operations, factoring in instruction-level parallelism using issue cycle metrics from calibration. Concurrency constraints assess how thread block dimensions influence warp residency, balancing register usage, shared memory allocation, and thread counts, while accounting for diminishing returns from parallel streaming multiprocessor (SM) activation, as observed in calibration data (Figure~\ref{fig:concurrency-scaling})~\cite{gao2022, nvidiaPtx2025}. By modeling these interactions, the framework enables developers to select configs that optimize performance and energy consumption without requiring extensive runtime testing.

\vspace{-1mm}
\paragraph{\textit{Dynamic Power Modeling}}
\label{subsec:power-model}
The power model estimates GPU energy consumption by decomposing dynamic power into functional unit activities and static leakage~\cite{Isci2003, hong_integrated_2010, leng_gpuwattch_2013} with architectural calibration and shape-aware corrections. Total power (\(P_{\text{total}}\)) is computed in Algorithm~\ref{alg:flipflop}, lines 16–35, using calibrated power coefficients (\(\beta_u\)) for functional units (\(\mathcal{U} = \{\text{FP32}, \text{INT}, \text{SFU}, \text{ALU}, \text{Mem}\}\)).

Refinements include:
\begin{itemize}
    \item \textit{Aspect ratio correction}: 
    Adjusts for thread block geometry effects on memory coalescing efficiency (Algorithm~\ref{alg:flipflop}, line 27 (A\ref{alg:flipflop}.L27 in short)), capturing the well-documented principle that consecutive threads accessing consecutive addresses maximize memory bandwidth~\cite{nvidiacudaguide2025}. 

    \item \textit{Transient power compensation}: Accounts for short-duration kernel effects using microbenchmark-derived ramp factors.
    \item \textit{Coalescing-aware DRAM power}: Penalizes non-contiguous memory accesses by adjusting base memory power (A\ref{alg:flipflop}.L28).
\end{itemize}

By separating out architectural parameters and measuring them through calibration, our model maintains portability while capturing power variance induced by thread block geometry changes.

\vspace{-1mm}
\paragraph{\textit{Energy Consumption Modeling}}
\label{subsec:energy-model}

The energy model integrates execution time and power predictions to estimate total energy consumption, validated against hardware measurements. Total energy (\(E_{\text{total}}\)) is calculated in A\ref{alg:flipflop}.L43, as \(E_{\text{total}} = T_{\text{exec}} \times (P_{\text{dyn}} + P_{\text{static}}) + E_{\text{overhead}}\), where \(T_{\text{exec}}\) is the predicted execution time, \(P_{\text{dyn}}\) is dynamic power, \(P_{\text{static}}\) is leakage power, and \(E_{\text{overhead}}\) accounts for kernel launch costs. The workflow involves three phases. Config space generation enumerates valid thread block dimensions (\eg{} 1$\times$32 to 1024$\times$1024), filtering out infeasible configs based on hardware constraints like warp alignment and resource constraints (\eg{} register and shared memory limits). Predictive analysis computes execution time and power using the respective models, combining them to estimate energy. Empirical validation executes configs using the Kernel Tuner framework~\cite{kerneltuner}, collecting actual energy measurements via NVML hardware counters and comparing them to predictions for model refinement. This process enables the identification of Pareto-optimal energy-performance configs, streamlining the development of sustainable GPU applications.
   
\vspace{-2mm}
\noindent
\paragraph{\textit{Configuration Space Exploration}}
\label{subsec:configexplore}
Our methodology explores the high-dimensional config space (\eg{} block\_x, block\_y, GPU power cap) to identify operating points that minimize energy while meeting performance constraints. Infeasible configs are filtered based on hardware limits (\eg{} occupancy).
 Following this constraint filtering, the method adapts the GPU’s operating frequency to the active power cap using a simplified voltage-frequency scaling relation. Specifically, we adjust the frequency via the equation:
$
f_{\text{adj}} \;=\; f_{\text{base}}\;\times\; \sqrt[K]{\frac{P_{\text{cap}}}{P_{\text{TDP}}}}
$
where \(f_{\text{base}}\) represents the GPU’s default clock, \(P_{\text{TDP}}\) is its thermal design power, \(P_{\text{cap}}\) is the desired power limit, and \(K\) is a fitting integer
(typically set to 3)~\cite{Amit2017dvfs, han2025dvfsawarednninferencegpus}. This formulation is motivated by standard dynamic voltage and frequency scaling (DVFS) principles—in processors, dynamic power is proportional to \(f \times V^2\)~\cite{Amit2017dvfs, han2025dvfsawarednninferencegpus} while the operating frequency f scales roughly linearly with voltage V~\cite{keller2015opportunities}. Consequently, if we assume an approximate cubic relationship (i.e., \(P \propto f^3\)), then reducing the power cap to a fraction of \(P_{\text{TDP}}\) necessitates a corresponding scaling down in frequency. A Pareto set \(\mathcal{P}\) is constructed to include configs that reduce energy while meeting performance targets (A\ref{alg:explore}.L8). This automated exploration eliminates manual tuning, providing developers with optimal configs for energy-efficient GPU applications across diverse workloads.

In summary,
this four-phase pipeline (Fig.~\ref{fig:workflow}),
systematically predicts how kernel block dimensions and power limits jointly influence execution time and energy consumption. Phase 1 inspects PTX-level instructions to establish memory and compute characteristics. Phase 2 tailors the model to the specific hardware via microbenchmark-derived latencies and power coefficients. Phase 3 merges these inputs into coupled time and power models, capturing concurrency effects. Finally, Phase 4 explores the config space—pruning infeasible points and retaining only Pareto-optimal solutions—to discover an energy-optimal \gpu{} kernel config.

\vspace{-1mm}
\noindent
\paragraph{\textbf{Results---RQ2}}
\label{sec:RQ3_results}

\begin{figure}[ht!]
    \centering
    \vspace{-2mm}
    \includegraphics[width=0.9\columnwidth]{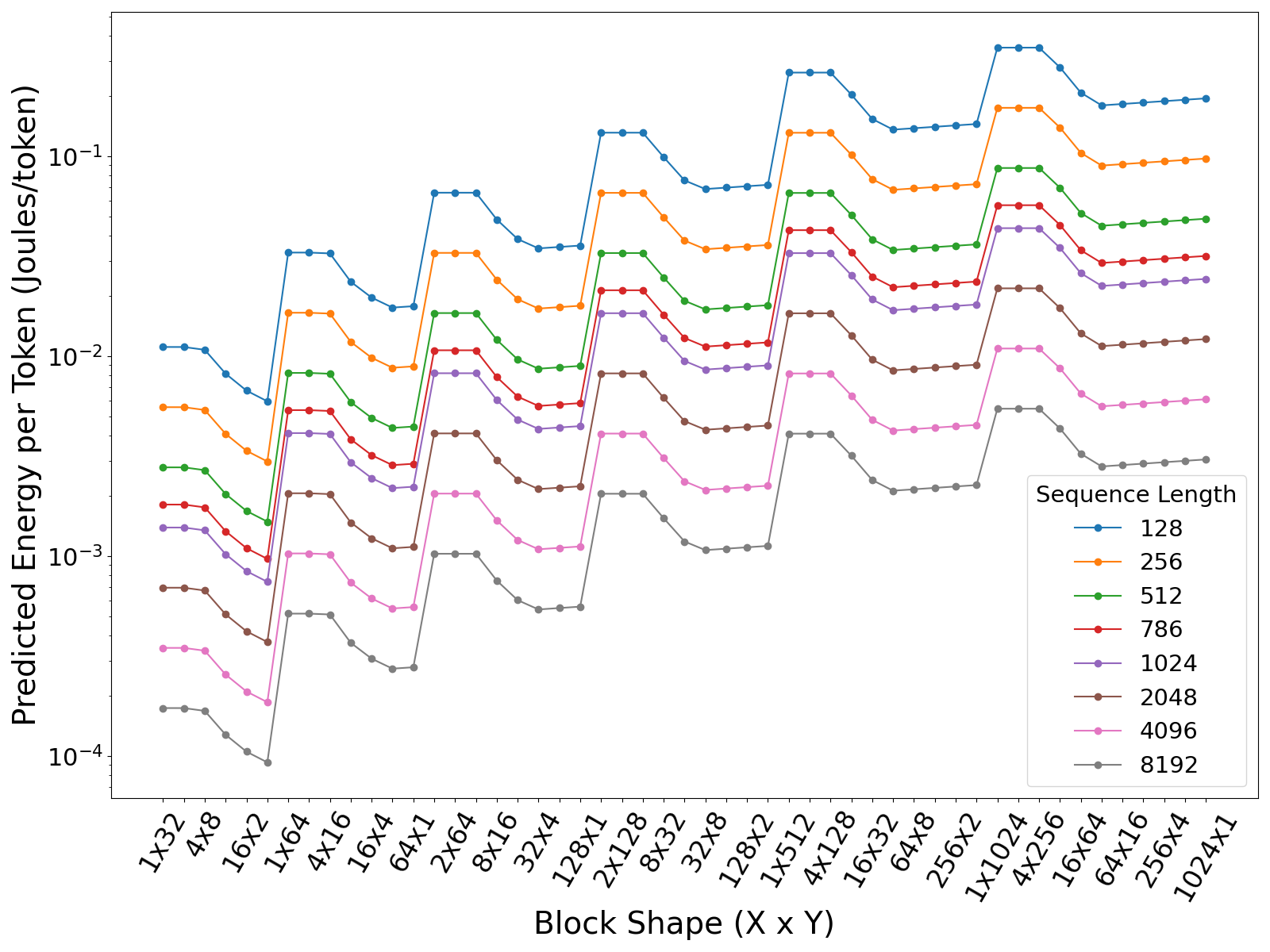}
    \vspace{-5mm}
    \caption{Predicted energy per token as a function of block config. Each line corresponds to a different seq. length}
    \label{fig:rq3-predicted-energy-blockdims}
    \vspace{-2mm}
\end{figure}

\begin{figure}[ht!]
    \centering
    \vspace{-2mm}
    \includegraphics[width=0.9\columnwidth]{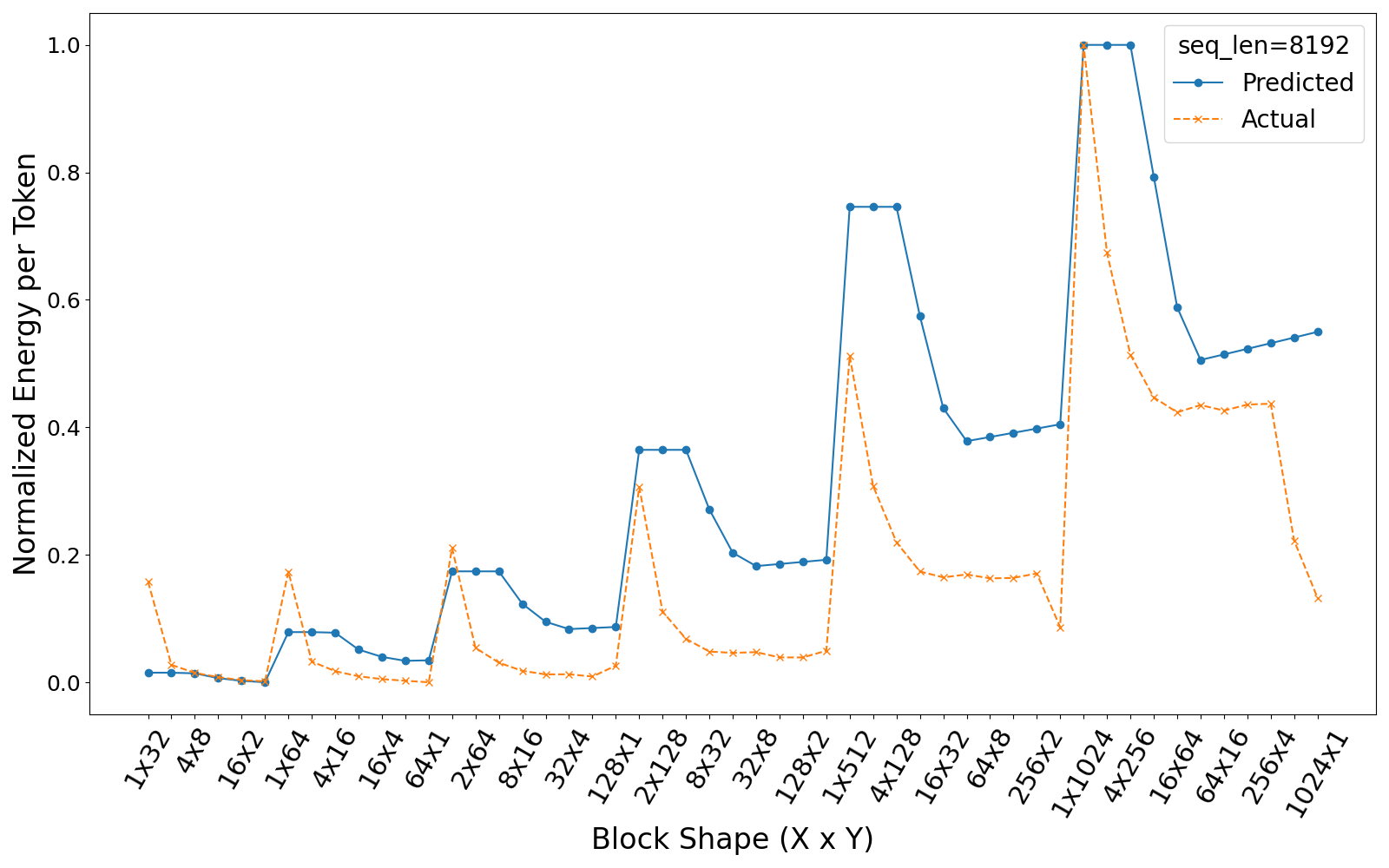}
    \vspace{-5mm}
    \caption{Trends between Actual and Predicted Energy}
    \label{fig:normalized_metrics}
    \vspace{-3mm}
\end{figure}

To evaluate \emph{FlipFlop}’s ability to predict and optimize energy consumption, we analyze its performance on the multi-head attention (MHA) kernel across sequence lengths 128--8192, 
validated on an RTX 5000 GPU and an additional RTX 3070 to ensure architectural portability. We also extend validation to diverse CUDA kernels (convolution, laplace3d, matMul, reduction, scalarProd, transpose, vecAdd) to confirm generalization across kernel types. 
As no existing approach in Table~\ref{tab:related-work-comparison} provides pre-execution static energy modeling for kernel configs, we establish a rigorous baseline using Kernel Tuner~\cite{kerneltuner,schoonhoven_going_2022} for exhaustive runtime profiling and actual hardware measurements.
Figure~\ref{fig:rq3-predicted-energy-blockdims} illustrates predicted energy per token for the MHA kernel across block configs, closely mirroring measured trends from RQ1 (Figure~\ref{fig:rq1-energy-blockdims}). For sequence length 8192, Figure~\ref{fig:normalized_metrics} shows normalized predicted and actual energy values, revealing consistent local minima and maxima with an accuracy of 83\% in identifying Pareto-optimal configs (maxima and minima) across all tested kernels.

\emph{FlipFlop}’s static PTX analysis, combined with a brief calibration phase, effectively identifies inefficient block shapes, reducing energy consumption up to 20\% on average compared to unconstrained/default Power, with less than 5\% performance loss. Across architectures, the model maintains prediction accuracy on the RTX 3070, consistent with RTX 5000 results, confirming portability.

Adaptive power capping, refined by real-time feedback, reduces worst-case prediction errors from 10--12\% to below 5\% by adjusting coalescing and latency parameters when power deviates by more than 10 W. The power limit is set as \(\texttt{power\_limit} = \min(\text{TDP}, \alpha \hat{P} + \beta \Delta P_{\text{history}})\), where \(\hat{P}\) is the predicted power from the static model and \(\Delta P_{\text{history}}\) is the cumulative deviation between predicted and actual power over recent iterations. This approach ensures robust energy savings under realistic constraints, with no single block config optimal across all workloads, underscoring the need for multi-objective optimization.

\textbf{Statistical validation of prediction accuracy.} To rigorously assess \emph{FlipFlop}'s predictive reliability, we conduct comprehensive rank correlation analysis using Spearman's $\rho$ (robust to non-linear relationships) and Kendall's $\tau$ (robust to outliers), following best practices for non-normally distributed data. Energy prediction achieves strong correlation across sequence lengths, \eg{} $\rho = 0.846$ ($p = 5.53 \times 10^{-15}$) at seq\_len=128, $\rho = 0.875$ ($p = 4.69 \times 10^{-17}$) at seq\_len=512, $\rho = 0.834$ ($p = 3.17 \times 10^{-14}$) at seq\_len=1024, with aggregate mean $\rho = 0.857 \pm 0.019$ and Kendall $\tau = 0.653 \pm 0.028$. Power modeling exhibits consistent accuracy ($\rho = 0.840 \pm 0.031$, all $p < 10^{-11}$) and execution time prediction achieves $\rho = 0.66 \pm 0.067$ (all $p < 10^{-6}$), validating architectural generalizability. Paired t-tests confirm statistically significant agreement ($t = 12.52$, $df = 50$, $p < 10^{-15}$). Translating correlation metrics into practical performance, \emph{FlipFlop}'s top-20 predictions capture 94\% of actual top-20 energy-efficient configs, demonstrating reliable optimization guidance without exhaustive profiling. 
\textit{Comparing against Bayesian optimization}: Bayesian optimization evaluated 800 configs yet missed optimum by 9.3\%, while \emph{FlipFlop} reaches optima after only 4.4 evaluations, demonstrating that static PTX analysis provides fundamental efficiency advantages over iterative search methods.

\textbf{Ablation study validating aspect ratio correction.} To validate the aspect ratio correction (Algorithm~\ref{alg:flipflop}, line 27), we compare model performance with and without this refinement. The correction penalizes configs deviating from memory-efficient access patterns, capturing documented GPU behavior where memory bandwidth is maximized when consecutive threads access consecutive addresses~\cite{nvidiacudaguide2025}. For MHA kernels, disabling the correction degrades energy prediction correlation by 83\% ($\rho$: 0.852 to 0.142), as the model loses ability to distinguish configs with poor energy efficiency due to bank conflicts and suboptimal warp scheduling.

\vspace{-1mm}
\paragraph{\textit{Answer to RQ2.}}
\emph{A hybrid static-and-dynamic model that combines PTX-level feature extraction can predict per-kernel energy usage to within 5--10\% of measured values and enable \textbf{adaptive power capping} strategies. This yields significant energy savings (20\% on average) while maintaining throughput above 90--95\% of peak.}

\vspace{-1.5mm}
\section{RQ3: Practical Impact of Optimization}
\label{sec:rq3}
\vspace{-1.5mm}
\noindent
\paragraph{\textbf{Methods---RQ3}}

We evaluate \emph{FlipFlop}’s effectiveness in reducing developer effort and computational resources by comparing its static optimization approach against exhaustive profiling of the MHA kernel across sequence lengths 128--8192, with a batch size of four, and 16 heads at 256 dimensions per head. We define 66 thread block configs per sequence length, where \texttt{block\_x} and \texttt{block\_y} are powers of two (1--1024), constrained by total threads (32--1024, divisible by 32) per NVIDIA warp-alignment rules, each profiled five times for statistical reliability.

Our proposed approach, \emph{FlipFlop}, uses static analysis to recommend only Pareto-optimal configs, minimizing profiling by recommending only high-performing setups, that balance energy and execution time, incorporating power limits (100--250W, 25W increments) to expand the search space to 528 configs. \emph{FlipFlop},
significantly reduces the number of configs profiled while preserving energy-efficient performance, saving substantial time, energy, and computational resources.

We employ the following metrics to quantify savings.

\begin{enumerate}
    \item \textbf{Configuration Reduction Ratio (CRR)}: Fraction of configs avoided, computed as
    $\text{CRR} = 1 - \frac{\text{\# \emph{FlipFlop}-recommended configs}}{\text{Total valid configs}}$

    \item \textbf{Energy and Carbon Savings}: Avoided profiling energy and associated emissions computed as
    $\Delta E = \sum (E_{\text{run}} \times N_{\text{runs}})$ and $\quad \Delta \text{CO}_2 = \Delta E \times \text{CI}_{\text{regional}}$
     where $E_{\text{run}}$ is measured energy per kernel execution and CI is carbon intensity ($0.384$ kg CO\textsubscript{2}/kWh for US~\cite{worldCarbonIntensity}. We choose US as it houses nearly half of the global datacenters~\cite{visualcapitalistRankedCountries}).

\end{enumerate}

\noindent
\paragraph{\textbf{Results---RQ3.}
\textit{Efficiency Gains}}
\emph{FlipFlop} significantly reduces the optimization search space by 93.4\%, recommending an avg. of 4.4 configs per seq. length out of the total 66 configs, as detailed in Table~\ref{tab:crr-results}. The reduction effectiveness varies by sequence length: at a sequence length of 128, it achieves a 98.5\% reduction with only 1 config, while at 4096, it offers a 95.5\% reduction with 3 configs, and at 8192, a 97.0\% reduction with 2 configs. This demonstrates the approach’s ability to adaptively minimize the search space across different workload sizes.

\begin{table}[ht!]
\centering
\vspace{-2mm}
\caption{Config Reduction Ratio (CRR) per sequence length}
\vspace{-3mm}
\label{tab:crr-results}
\setlength{\tabcolsep}{2.5pt}
\renewcommand{\arraystretch}{1}
\resizebox{\columnwidth}{!}{
\scriptsize
\begin{tabular}{l|rrrrrrrr|r}
\textbf{Seq. Length} & 128 & 256 & 512 & 786 & 1024 & 2048 & 4096 & 8192 & \textbf{Avg.} \\
\hline
\textbf{Total Configs} & 66 & 66 & 66 & 66 & 66 & 66 & 66 & 66 & 66 \\
\textbf{Recommended} & 1 & 5 & 6 & 6 & 6 & 6 & 3 & 2 & \textbf{4.4} \\
\textbf{CRR} & 0.985 & 0.924 & 0.909 & 0.909 & 0.909 & 0.909 & 0.955 & 0.970 & \textbf{0.934} \\
\end{tabular}
}
\vspace{-3mm}
\end{table}

\paragraph{\textit{Resource Savings}}
The configuration reduction translates into substantial resource conservation, saving 14.3 minutes of computation time, which is 172,986× faster than dynamic tuning. This efficiency also yields energy savings of 188,680 J, equivalent to 52.4 watt-hours, sufficient to charge three smartphones~\cite{epaGreenhouseEquivalencies}.  For context, using US coal-fired power generation as a reference point (based on EPA equivalency metrics~\cite{epaGreenhouseEquivalencies}), this energy savings corresponds to approx. 19.4g CO\textsubscript{2}e~\cite{epaGreenhouseEquivalencies} avoided, though actual emissions may vary by regional power mix and generation source. 

\vspace{-1mm}
\paragraph{\textit{Production Context for Profiling Costs}}

While FlipFlop's config space reduction saves 14.3 minutes and 188,680 J per optimization session, the true value emerges in production deployment contexts. This one-time optimization cost amortizes after approximately 150 inference runs (at 1,257 J average per run for seq\_len=512). 
More critically, AI is projected to use 165–326 terawatt-hours annually by 2028, enough to power 22\% of US households~\cite{mittechreview2025ai}.
Even 1\% energy inefficiency from suboptimal configs—running continuously in production—generates ongoing waste far exceeding any profiling overhead. 
Unlike iterative profiling methods that risk convergence failure or require re-tuning when config spaces change, FlipFlop's static analysis provides deterministic near-optimal configs without heavy kernel execution. This eliminates both the initial profiling cost and the compounding production inefficiency that runtime methods may introduce. Furthermore, when compared against non-exhaustive alternatives like Bayesian optimization (which evaluated 785 configs yet missed the optimum by 9.3\%, Section~\ref{sec:RQ2}), FlipFlop's 392× efficiency advantage stems not from avoiding exhaustive search but from eliminating runtime profiling altogether—a fundamental difference from tools like KLARAPTOR~\cite{brandt_klaraptor_2019} and Kernel Tuner~\cite{kerneltuner} that require kernel execution even with smart search strategies.

\vspace{-1mm}
\paragraph{\textit{Developer Impact}}
\emph{FlipFlop} transforms kernel configuration workflows by eliminating manual tuning effort, reducing configs by 93.4\% across sequence lengths. This cuts experimentation time by 15.1×, shrinking hours to minutes, and reduces optimization overhead by 95.6\% in energy consumption, enabling sustainable development. 
Its efficiency scales to complex models, allowing 100-kernel optimizations in under 12 minutes.

\vspace{-1mm}
\paragraph{\textit{Scaling Analysis}}
The resource savings demonstrated by static optimization compound substantially as problem complexity increases. Experimental projections show that optimizing a 100-kernel model would require approx. 3 hours with dynamic runtime profiling versus just minutes with \emph{FlipFlop}. This scaling efficiency stems from \emph{FlipFlop}'s fundamental advantage: its static analysis methodology remains computationally efficient regardless of problem scale. As context windows expand to 1M+ tokens requiring optimization across hundreds of sequence lengths, or when tuning entire models containing dozens of diverse kernels, the relative time and energy savings per kernel grow linearly with the number of configs evaluated. For example, extending our approach to 100 sequence lengths would yield 150× time savings. This scalability ensures that the environmental benefits and developer time savings become increasingly significant when applied to complex modern architectures with massive context windows and heterogeneous kernel collections, making \emph{FlipFlop} particularly valuable for next-generation LLMs where iterative tuning would otherwise incur prohibitive computational costs.
\vspace{-1mm}
\paragraph{\textit{Answer to RQ3.}}
\emph{FlipFlop reduces developer effort by 93.4\%, accelerates optimization by 172,986×, and saves \textbf{188,680J} (19.4g CO\textsubscript{2}e) per workload. This transforms kernel tuning from resource-intensive profiling to efficient static analysis—enabling practical iterative optimization while reducing environmental impact.}

\vspace{-2mm}
\section{Case Study: Energy Optimization in LLM Inference}
\label{sec:case-study}

\textbf{Scenario and Motivation.} This case study explores how a software engineer implements CodeLlama 7B, an open-source code model, using low-level CUDA programming rather than high-level frameworks such as PyTorch. The engineer's primary motivation is maximizing performance and energy efficiency for edge devices with limited computational resources. 
By implementing core operations directly in CUDA, they achieve fine-grained control over memory access patterns, precise resource management, elimination of framework overhead, and optimal energy-per-inference metrics.

\vspace{1mm}
\noindent
\textbf{Kernel Optimization Challenge.} They focus on optimizing the MHA kernel for incrementally growing context windows, as it represents the most critical optimization target due to its substantial compute share in transformer models. Traditional tuning approaches---brute-force parameter search or runtime heuristics---prove problematic due to exponential config space growth, hardware-specific performance characteristics, repeated tuning requirements for new architectures, and prohibitive energy costs during exploration. These limitations motivate the engineer to adopt FlipFlop.

\vspace{1mm}
\label{subsec:methodology}
\noindent
\textbf{Methodology.} The developer begins with static analysis using FlipFlop to identify promising thread configs from the vast combinatorial space, narrowing candidates to a tractable subset for empirical validation. Using HumanEval's 164 programming problems as realistic workloads, six representative thread block configs are selected from FlipFlop's recommendations: 1024×1 (1D arrangement), 32×32 (wide config), 256×4 (2D narrow pattern), 512×1 (reduced 1D layout), 2048×1 (expanded 1D structure), and 64×16 (balanced geometry), systematically exploring thread parallelism, memory access patterns, and resource utilization trade-offs.
FlipFlop's automated workflow generates specialized CUDA binaries for each config, creating temporary source files with config-specific constants, compiling distinct executables using \textit{nvcc}, organizing outputs into timestamped directories for reproducibility, and validating functionality through prompt-based smoke tests. During benchmark execution, the system samples GPU power at 50ms intervals, records per-prompt latency, calculates energy through numerical integration, and logs latency (seconds), average power (watts), and total energy (joules). The experimental protocol implements best practices: sequential HumanEval executions avoid GPU resource contention, comprehensive process cleanup eliminates residual state, user-specific management ensures isolation, and timeout mechanisms handle failures. Each config processes all 164 problems, generating 26,256 individual measurements across three metrics per prompt per config.

\vspace{1mm}
\noindent
\textbf{Results \& Discussion.}
The custom config (64×16) recommended by FlipFlop demonstrated superior energy efficiency, consuming 15.7\% less energy (5510.7J vs 6536.0J) while achieving 14.0\% lower latency (27.97s vs 32.51s) compared to the default config. This improvement stems from the 2D thread arrangement better aligning with the kernel's memory access patterns, reducing shared memory bank conflicts while improving coalesced global memory accesses.
The energy-performance tradeoffs reveal important architectural insights. While power draw shows moderate variation (195.6-202.1W), latency differences drive significant energy variance due to E = P × t. The 64×16 config's 14\% latency reduction directly enabled proportional energy savings, confirming execution time optimization as critical for efficiency. Notably, the 2048×1 config consumed 17.6\% more energy than our recommendation despite similar latency to default, demonstrating how improper thread sizing triggers suboptimal power scaling through resource contention. 

This case study demonstrates that energy-aware kernel optimization delivers concrete engineering advantages. Three system-level implications emerge: (1) Static analysis reduces optimization effort by orders of magnitude, narrowing thousands of configs to six viable candidates while identifying global optima; (2) Significant energy savings are achievable solely through thread reshaping without algorithmic modifications; (3) Future GPUs with tightening thermal limits may prioritize rapid completion over momentary power reduction, validating FlipFlop's energy modeling approach.

\vspace{-2mm}
\section{Implications}
\label{sec:implications}

The \emph{FlipFlop} framework enhances energy-efficient computing by using static analysis to optimize GPU kernel configs, embedding energy efficiency as a core design principle.
\textbf{Software Developers:} \emph{FlipFlop} empowers developers to create energy-efficient GPU applications without requiring deep hardware expertise, embedding energy optimization directly into the development process. By analyzing PTX code, \emph{FlipFlop} enables developers to identify optimal kernel configs early. As demonstrated in Section~\ref{sec:case-study}, \emph{FlipFlop} accelerates discovery of energy-optimal configs, ensuring developers---regardless of hardware knowledge---can build efficient software pipelines that scale effectively in deployment with minimal performance overhead.
\textbf{System Architects:} \emph{FlipFlop}'s ability to predict optimal configs without runtime profiling reduces complexity in designing portable, adaptive systems. By offering explainable guidance on resource allocation, \emph{FlipFlop} enables architects to balance performance and energy efficiency, fostering resilient systems that adapt to evolving hardware and workload demands.
\textbf{Hardware Vendors:} \emph{FlipFlop}'s insights into memory access patterns and power dynamics inform optimizations in GPU architectures and tools such as CUDA~\cite{nvidiaCuda2025} and ROCm~\cite{rocm2025}. Integrating its predictive models into development kits enhances access to energy-efficient features.
\textbf{Sustainability Researchers:} By enabling proactive energy optimization without runtime costs, \emph{FlipFlop} supports development of eco-friendly systems for AI, scientific simulations, and beyond. Its approach aligns with global efforts to minimize computing's carbon footprint, providing a replicable model for studying energy-performance trade-offs and fostering sustainable design principles across industries.

\vspace{-2mm}
\section{Threats to Validity}
\label{sec:threats}

\textbf{Internal validity} risks arise from model assumptions (e.g., memory coalescing, thread block impacts) that may not hold across all GPUs or workloads, potentially skewing predictions. We mitigate this with a hybrid static-dynamic model, reducing prediction errors to below 5\% via corrections. Calibration on the RTX 5000, with additional validation on an RTX 3070, ensures hardware-specific accuracy. Energy measurement inaccuracies are minimized by using NVIDIA’s NVML~\cite{nvidiaNVIDIAManagement} and CUPTI~\cite{nvidiaCUPTIx2014} tools
and following best practices from prior work~\cite{RajputFecom2024}. \textbf{External validity} is challenged by testing primarily on NVIDIA GPUs, which may limit generalizability to other architectures. We address this with an architecture-agnostic model validated on both RTX 5000 and RTX 3070 GPUs. 
Similarly, focusing on MHA kernels may not extend to all workloads, so we evaluate additional CUDA kernels (e.g., convolution, matrix multiplication, reduction) to confirm 83\% accuracy in identifying Pareto-optimal configs across diverse kernel types (Section~\ref{sec:RQ2}).

\vspace{-2mm}
\section*{Conclusions and Future Work}
\label{sec:conclusion}

We introduce FlipFlop, a framework providing optimized GPU kernel configurations calibrated to underlying hardware through PTX-level feature extraction via static analysis and performance-power modeling. By systematically modeling memory access patterns, arithmetic intensity, and concurrency limits, our framework reveals how moderate thread block configurations can better balance utilization and power draw than occupancy-focused heuristics. Our evaluations on LLM workloads involving MHA confirm that jointly tuning kernel shapes and GPU power caps yields meaningful energy reductions without sacrificing throughput or model quality.

Looking ahead, we plan to expand the framework's applicability to broader AI and scientific computing kernels and investigate more granular power-capping strategies based on dynamic workload phases. Incorporating dynamic voltage-frequency scaling (DVFS) within our static modeling will further refine energy-saving strategies under rapidly changing load conditions typical in real-world large-scale LLM inference pipelines. Our controlled evaluation uses single-kernel execution to isolate configuration effects, reflecting common deployment patterns (dedicated GPUs, containerization) that minimize contention for predictable latency. While our calibration models resource-sharing dynamics through power-law SM scaling and bandwidth saturation, comprehensive validation under heavy multi-stream contention represents valuable future work. We also plan to develop libraries abstracting low-level optimizations---thread block tuning and power capping---while preserving developer control for fine-grained customization. These advancements will further streamline sustainable GPU software development, supporting efficient, large-scale LLM inference with convenience across diverse environments.

\bibliographystyle{ACM-Reference-Format}

\bibliography{reference}
\balance

\end{document}